# Demonstration of variable angle Super-Heterodyne Dynamic Light Scattering for measuring colloidal dynamics

Daniel Crowley[a,⊗], Riande I. Dekker[a,b,c,⊗], Denis Botin[a], Nicole Schärtl[a,d], Annalena Groß[a,d], Aakanksha Agarwal[a,e], Sabrina Heidt[a], Jennifer Wenzl[a,f], Negar Zaghi[a,g], Evgenii Vorobev[a], and Thomas Palberg[a§]

[a] *Institut für Physik, Johannes Gutenberg Universität, 55128 Mainz, Germany*

[b] *Van der Waals-Zeeman Institute, Institute of Physics, University of Amsterdam, Science Park 904, 1098 XH Amsterdam, The Netherlands*

[c] *Van 't Hoff Laboratory of Physical and Colloid Chemistry, Debye Institute for Nanomaterials Science, Utrecht University, Padualaan 8, 3584 CH, Utrecht, The Netherlands.*

[d] *Institut für Physikalische Chemie, Albert-Ludwigs-Universität Freiburg, 79104 Freiburg, Germany*

[e] *Birla Institute of Technology and Science, Pilani Vidya Vihar, Pilani, Rajasthan 333031, India*

[f] *GSI Helmholtzzentrum für Schwerionenforschung, 64291 Darmstadt, Germany*

[g] *Dept. of Chemical and Biological Engineering, University of British Columbia, Vancouver, BC V6T 1Z4, Canada.*

[⊗] *both authors contributed equally*

[§] *correspondence should be sent to palberg@uni-mainz.de*

**We demonstrate a prototype light scattering instrument combining a frequency domain approach to the intermediate scattering function from Super-Heterodyning Doppler Velocimetry with the versatility of a standard homodyne Dynamic Light Scattering goniometer setup for investigations over a large range of scattering vectors. Comparing to reference experiments in correlation-time space, we show that the novel approach can determine diffusion constants and hence hydrodynamic radii with high precision and accuracy. Possible future applications are discussed shortly.**

## Acknowledgements

We like to thank G. Nägele and S. Egelhaaf for their continued interest in our approach resulting in many fruitful discussions. We thank R. Sreij (Univ. Bielefeld) for providing SAXS particle size measurements. Financial Support of the Deutsche Forschungsgemeinschaft (DFG, Grant no. Pa459/18-2) and the Inneruniversitäre Forschungsförderung der Johannes Gutenberg Universität, Mainz is gratefully acknowledged. E. Vorobev and N. Zaghi were trainees within the IAESTE program and received fellowships by the Deutsche Akademische Austauschdienst (DAAD). R. I. Dekker was an Erasmus+ Trainee and recipient of a fellowship through the Duitsland Intstituut Amsterdam (DIA). S. Heidt is a fellow of the Graduate School of Excellence MAINZ and received funding by the DFG (Grant DFG/GSC 266).

# Introduction

Structured fluids offer a wide range of interesting fundamental problems and technological applications [1]. Colloidal suspensions in particular consist of (spherical) solid particles suspended in a carrier liquid. Being able to form colloidal fluids, glasses and crystals, they have proven to be valuable mesoscopic models for a large variety of fundamental problems of Statistical Mechanics and Condensed Matter Physics. These range from the very existence of atoms [2], over studies on gas-liquid interfaces [3], to matter in external fields [4], and to active matter [5]. In many instances, such systems pose considerable experimental challenges like meta-stability, multiple relaxation times, non-ergodicity or a pronounced turbidity.

The latter issue creates particularly demanding difficulties for any experiment employing photon correlation spectroscopy (PCS) or dynamic light scattering (DLS), which is the standard technique to obtain colloidal dynamics [6, 7, 8]. More specifically, it concerns effects due to the difference in refractive index of particles and solvent. This is also known for W/O emulsions [9] or suspensions in organic media [10] but much more pronounced in aqueous suspensions. It leads to extinction as well as to multiple scattering and strongly distorts measured correlation functions [11, 12]. In water, only for systems of low refractive index, index-matching is feasible, e.g. for perfluorinated polymer latex spheres [13]. However, full dynamic information can also be obtained for weakly or moderately multiple scattering systems from cross-correlation instruments [14, 15, 16, 17], some of which are even available commercially [18]. A second issue is the non-egodicity of samples displaying slow structural relaxations like polycrystalline solids, glasses or gels. Here, time averages do not coincide with ensemble averages. This forbids a straightforward use of the Siegert relation $\hat{g}_I(q,t) = 1 + [\hat{g}_E(q,t)]^2$ to relate the measurable normalized intensity auto-correlation function $\hat{g}_I(q,t))$ to the desired normalized field auto-correlation function $\hat{g}_E(q,t)$ [19]. $\hat{g}_E(q,t)$, also known as intermediate scattering function or dynamic structure factor, is the Fourier-transform of the van Hove space-time correlation function describing the dynamics in real space. In DLS, typical diameters of the detection volume are restricted to some 50 - 100µm to preserve coherent illumination. Thus, a DLS measurement records time averaged data from small ensembles. To obtain $\hat{g}_E(q,t)$ in slow systems, additional measures must be taken to ensure correct ensemble averaging [20]. Several procedures have been reported for that step [21, 22, 23, 24, 25]. The most recent ones even simultaneously address multiple scattering and non-ergodicity, and they return ensemble averaged single-scattering dynamic structure factors [26]. However, all of these techniques still rely on the Siegert-relation and do not supply direct access to $\hat{g}_E(q,t)$, In addition, they often demand some sophisticated optical and mechanical instrumentation.

We here report on a prototype light scattering instrument combining the versatility of a goniometer setup for investigations over a large range of scattering vectors with the frequency domain approach of Super Heterodyne Dynamic Light Scattering (SH-DLS) [27]. Analysis in frequency space was introduced already early for homodyne (self-beating) light scattering experiments to study diffusive properties of biological macro-molecules [28] but not much followed after the rapid development of time domain DLS. By contrast, the optical mixing technique of heterodyning (beating of scattered light with a local oscillator) works very well in frequency space [29] and became a standard technique in flow or Doppler velocimetry [30]. Super-heterodyning (SH), in addition, allows separating any homodyne contribution and low frequency noise by adding an additional frequency shift, $\omega_{SH}$, between scattered light and local oscillator [31]. Integral measurements collect scattered light from a large observation volume or even the complete cross section of the sample cell and thus provide an excellent ensemble average [32]. The heterodyne part of the super-heterodyne spectra contains all the relevant information in terms of the Fourier transform of the intermediate scattering function [33]. A small angle configuration is optimally suited for Doppler velocimetric investigations of flow and self-diffusion irrespective of sample structure [34, 35]. Very recently, we also implemented a facile way of correcting for multiple scattering in small angle scattering [27]. This allowed detecting dynamics in turbid samples with a transmission as low as 20%, which is close to the detection limit of typical cross correlation experiments [15, 26] or heterodyne near field scattering (HNFS [36]). Modulated 3D cross correlation was shown to push that limit even further to values around 1% [18].

As one main application of DLS is particle sizing, a more informative quantity would be the limiting transmission for obtaining relaxation times with a certain precision (discussed e.g. in [37, 38] and references therein). An important benchmark for SH-DLS is the limit at which the average diffusion coefficient (and thus particle size) can still be determined with a statistical uncertainty of $\sigma \leq \Delta D_0/D_0 \approx 0.02$ corresponding to the accepted statistical uncertainty in homodyne time domain DLS particle sizing [39, 40]. This was possible in small angle SH-DLS at transmissions of 40% and larger [27]. With this approach we studied diffusion in turbid systems including flowing suspensions, active matter, and systems undergoing phase transitions [27, 41, 42]. For the competing cross correlation techniques, similar values have been reported, with modulated 3D cross correlation showing somewhat better marks due to the modulation induced increase of the intercept [18]. The limits for determining both mean particle size and width of a size distribution in poly-disperse samples have not been explored in a systematic way in the literature. It has, however, been shown that cross correlation techniques are excellently suited to extract multiple scattering free form factors as well as static and dynamic structure factors from cross correlation DLS data [18]. Cross correlation may, in addition, also give some limited access to information on advective and turbulent dynamics [43, 44]. These features and in particular the applicability of frequency domain analysis to the study of the length scale dependent diffusive dynamics in ordered and/or non-ergodic samples remain to be demonstrated for SH-DLS.

In what follows, we go a first important step beyond our previous work and introduce the first prototype instrument of a variable angle integral Super Heterodyne Dynamic Light Scattering (SH-DLS) instrument. This new instrument allows to cover the full range of scattering vectors known from goniometer-based DLS. We provide proof of principle for simple diffusion measurements on non-interacting, dilute suspensions at scattering angles between 15° and 135° from comparing to homodyne time domain DLS performed on the very same samples. We anticipate that SH-DLS may become a facile and versatile approach also to more complex situations including size dispersity analysis, non-ergodic materials and multiple scattering systems.

## Materials and Methods

**Multiple angle SH-DLS**

Fig. 1 shows a top view sketch of the new prototype experiment. Light of a HeNe-laser ($\lambda = 633$ nm) adjusted to propagate at sample height by mirrors (M) and passes a beam cleaner to remove stray light halos (1f arrangement of two lenses (L) with a circular aperture (CA) at focus. A beam splitter (BS) distributes the incoming light into illumination and reference beam paths. In each arm the light is frequency shifted by an acousto-optical modulator (BC, Bragg cell Model 3080-125, Crystal Technology Inc., US; frequency shift controller DFD 80, APE GmbH, Germany) after which the first order diffracted light is selected by another CA, before it is fed into single mode optical fibres (OF, OZ Optics Ltd., Carp, ON, Canada) with integrated fibre couplers (FC). Typical frequency differences $\omega_{BR}$ range between 1 and 8 kHz. The goniometer is a double-arm construction allowing to adjust the angle $\Theta$ between spatially fixed illuminating beam ($I_{ill}$) and reference beam ($I_{ref}$) by rotating the ring-shaped platform on which the reference beam sending and collecting optics are mounted. In this prototype, we still adjust $\Theta$ manually with a nonius reading precision of 0.02°, but integration of a motorized and piezo-controlled drive at a later stage is anticipated. The mechanical design allows both positive and negative angles. For the illumination beam sending optics ($S_I$), the lens (L) and a polarizer (P) are adjusted past the rear FC to obtain a parallel, vertically (V) polarized illuminating beam of 1-2 mm diameter within the sample cell (C) placed in the centre of the index match bath (IMB) filled with fused silica matching liquid (50350, Cargille Laboratories, Inc., France). Depending on beam diameter, the impinging integrated power of 5.1mW (determined at the outlet of $S_I$ and $S_R$) results in 0.4 to 1.6 mW/mm$^2$ within the sample. The same power is used for measurements at all angles. After passing the sample and the IMB the remaining light is stopped by a beam stop (B).

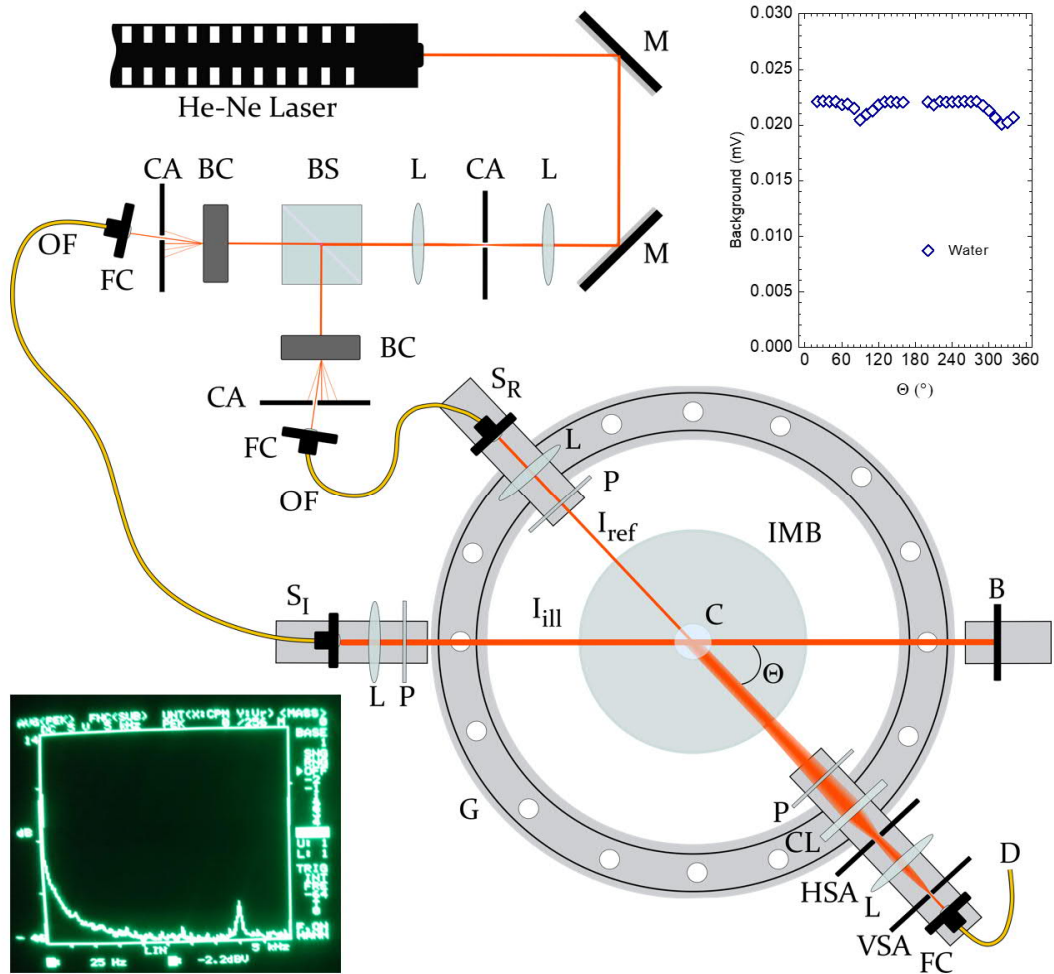

**Fig. 1** Top view sketch of the double arm goniometer with optical and mechanical components. (for details see text). Top right inset: Background intensity reading obtained from an experiment on pure water. Bottom left inset: historical smartphone shot of the first-light signal at the FFT-analyzer showing homodyne and heterodyne signal components.

The reference beam is treated similarly, albeit the sending optics (S$_R$) are adjusted to result in a smaller beam diameter of 0.5 mm. The detection side is aligned co-linear with the reference beam. A polarizer assures V/V detection and a horizontal slit aperture (HSA) rejects light received from out of plane directions. A combination of a cylindrical lens (CL) and a spherical lens (L), focus the light in the plane of a precision vertical slit aperture (VSA, Newport) which defines the detected scattering vector *q*, by rejecting all light focused off axis. The distance between the fiber coupler grin lens and the VSA defines the maximal extension of the observation volume. We collect singly scattered light emerging off the region illuminated by the illumination beam as well as multiply scattered light from within the complete observation volume. Of both contributions, the photons propagating exactly in the direction of the reference beam are selectively fed into another fiber leading to the detector (Hamamatsu H5783, SN:831-5833). A crucial point in the alignment of the double illumination path is avoidance of parasitic reflections under any detection angle. The top inset of Fig. 1 demonstrates the absence of reflections over the complete range of angles (with the exception of angles close to 0° and 180° due to interference with the illumination optics mounted there). The slightly lower reading (dips), here obtained around angles of 100° and 340°, can be shifted to other angles at constant angular difference between the dips by changing the orientation of the sample vial. It is therefore attributed to slight deviations from perfect circularity of the vial. It does not interfere with dynamic measurements but has to be accounted for if static data are requested. On the detector, the reference beam (acting as the local oscillator) and the scattered light superimpose, which gives rise to beats in the intensity observed at the detector. These are analyzed by a four channel Fast Fourier Transform analyzer (Ono Sokki D-3200, Compumess, Germany) to yield the power spectrum as a function of frequency, $f = \omega / 2\pi$. A smartphone snap shot of our first spectrum obtained with the new prototype is shown in the lower inset of Fig. 1. Note the broad homodyne signal extending from the origin and the smaller, but clearly discriminated peaked SH-DLS signal at 4 kHz corresponding to the frequency difference $\omega_{SH}$ between reference and illuminating beam. To obtain a noise level below 0.1 typically some 200 subsequent spectra are averaged corresponding to some five minutes measurement time. The noise level can be lowered to $10^{-3}$ by increasing the measurement duration accordingly. For particle characterization and as reference homodyne DLS-experiment we also employed a custom made multi-purpose instrument, described in detail elsewhere [45]. Evaluation of DLS data followed [38]. Initially, we assumed a constant ambient temperature to apply for all measurements. Since the instruments are located in differently climatised labs, we found it necessary to further note also the evolution of the ambient temperatures separately during each SH-DLS and DLS measurement series.

**Samples and sample conditioning**

Our sample for diffusion measurements in dilute aqueous suspension (lab code PnBAPS80) consisted of 35:65 W/W copolymer particles of Poly-n-Butylacrylamide (PnBA) and Polystyrene (PS), kindly provided by BASF, Ludwigshafen. Their nominal diameter and standard error based relative size dispersity index, PI, are given by the manufacturer as $2a_{nom} = 80.5$ nm (DLS) and PI = 0.19 (Hydrodynamic Chromatography), respectively. Our own form factor measurements using SAXS yield $2a_{SAXS} = 86.9$ nm and PI ≈ 0.08. The effective charges for PnBAPS80 are $Z_{eff,G} = 365.1 \pm 2.3$ (via TRS [45]) and $Z_{eff,\sigma} = 513 \pm 3$ (from conductivity[46]). Samples were prepared using batch conditioning procedures to first obtain a thoroughly deionized sample of number density $n = 5.1 \cdot 10^{18}$ m$^{-3}$ (from Debye Scherrer type Static Light Scattering) [47]. This was then further diluted with Milli-Q water to adjust the number density to $n = 1.2 \; 10^{17}$ m$^{-3}$ (volume fraction $\Phi = 0.0018$) and refilled into freshly cleaned and dried vials (Supelco, Bellefonte, PA, USA) without ion exchange resin added, capped to avoid contamination with dust, but not tightly sealed to allow equilibration with the CO$_2$ of ambient air. This results in so-called realistically salt free conditions. At $n = 1.2 \; 10^{17}$ m$^{-3}$ in the presence of $5.7 \; 10^{-6}$ mol/L of carbonate ions, the system is non-interacting and takes an isotropic gas-like structure (as corroborated by Static Light Scattering showing a static structure factor of $S(q) = 1 \pm 0.02$ for angles $15° \leq \Theta \leq 165°$).

**Data processing and evaluation.**

Our theoretical frame for single scattering is based on earlier work on homo- and heterodyning techniques in dynamic light scattering [6, 7, 48]. A theory of conventional heterodyne LDV using an integral reference beam set-up was outlined in [29]. Super heterodyne theory for integral measurements at low angles has been detailed in [33] and extended in [27] to include moderate multiple scattering. We therefore here only recall some basics relevant for the present experiments. The (single) scattering vector $\mathbf{q}_1 = \mathbf{k}_i - \mathbf{k}_1$ (where $\mathbf{k}_i$ and $\mathbf{k}_1$ are the wave vectors of the illuminating and one-time scattered light) is proportional to the momentum transfer from the photon to the scattering particle. Its modulus is given by $q = (4\pi\nu_S / \lambda_0) \sin(\Theta/2)$, where $\nu_S$ is the index of refraction of the solvent and $\lambda_0 = 633$ nm. The power or Doppler spectrum $C_{shet}(\mathbf{q},\omega)$, is the time Fourier transformation of the mixed-field intensity autocorrelation function, $C_{shet}(\mathbf{q},\tau)$:

$$C_{shet}(\mathbf{q},\omega) = \frac{1}{2\pi}\int_{-\infty}^{\infty} d\tau \; \exp(i\omega\tau)\, C_{shet}(\mathbf{q},\tau) = \frac{1}{\pi}\int_{0}^{\infty} d\tau \cos(\omega\tau)\, C_{shet}(\mathbf{q},\tau) \quad (1)$$

with circular frequency $\omega$ and correlation time $\tau$. For a scattered light field of Gaussian statistics the mixed field intensity autocorrelation function reads::

$$C_{shet}(\mathbf{q},\tau) = \left(I_r + \langle I_1(\mathbf{q})\rangle\right)^2 + 2I_r \langle I_1(\mathbf{q})\rangle \mathrm{Re}\left[\hat{g}_E(\mathbf{q},\tau)\exp(-i\omega_B\tau)\right] + \langle I_1(\mathbf{q})\rangle^2 |\hat{g}_E(\mathbf{q},\tau)|^2 \quad (2).$$

Here, $I_r$ is the reference beam intensity, and $<I_1(\mathbf{q})>$ is the time-averaged singly scattered intensity, and $\hat{g}_E(\mathbf{q},\tau) = g_E(\mathbf{q},\tau) / <I_1(\mathbf{q})>$ is the normalized field autocorrelation function. For homogeneous suspensions of interaction monodisperse, optically homogeneous, mono-sized spherical particles, $<I_1(\mathbf{q})>$ factorizes as:

$$\langle I_1(\mathbf{q})\rangle = I_0 n b^2(0) P(q) S(\mathbf{q}) \quad (3)$$

Here, $I_0$ is a constant comprising experimental boundary conditions like illuminating intensity, distance from the sample to the detector, and polarization details. $b^2(0)$ is the single particle forward

[8]scattering cross section, *n* is the particle number density, $P(q) = b(q)^2 / b(0)^2$ denotes the particle form factor and $S(\mathbf{q})$ the static structure factor [6, 7, 47]. For isotropic, fluid-like ordered samples, $S(\mathbf{q}) = S(q)$ and $\langle I_1(\mathbf{q}) \rangle = \langle I_1(q) \rangle$. In the present case, the suspension is in addition adjusted to be non-interacting and thus $S(q) = 1$. We assume the particles to undergo only Brownian motion with a single effective diffusion coefficient $D_{eff}$. For this simple case, the field correlation function is $\hat{g}_E(q,\tau) = \exp(-D_{eff} q^2 |\tau|)$. The power spectrum reads:

$$C_{shet}(q,\omega) = \left[I_r + \langle I_1(q) \rangle\right]^2 \delta(\omega)$$
$$+ \frac{I_r \langle I_1(q) \rangle}{\pi} \left[ \frac{q^2 D_{eff}(q)}{(\omega+\omega_B)^2 + (q^2 D_{eff}(q))^2} + \frac{q^2 D_{eff}(q)}{(\omega-\omega_B)^2 + (q^2 D_{eff}(q))^2} \right] \quad (4)$$
$$+ \frac{\langle I_1(q) \rangle^2}{\pi} \frac{2q^2 D_{eff}(q)}{\omega^2 + (2q^2 D_{eff}(q))^2}$$

For the equivalent expression in the presence of an additional drift velocity and of multiply scattered light, see [27]. $D_{eff}(q)$ depends on the probed length scale, in case collective diffusion is measured [49]. In the present case of unstructured suspensions, we expect collective and self- diffusion to coincide and hence to show no dependence on length scale. In fact, apart from hydrodynamic corrections [50], we expect to measure the free diffusion Stokes-Einstein-Sutherland Diffusion coefficient: $D_{eff} = D_0 = k_B T / 6\pi\eta a_h$. Here, $k_B$ is Boltzmann's constant, $\eta(T)$ the temperature dependent viscosity of water [51], and $a_h$ is the hydrodynamic radius).

The spectrum in Eqn. (4) contains three contributions: a trivial constant term centered at zero frequency, two super-heterodyne Lorentzians of spectral width $q^2 D_{eff}$ shifted away from the origin by the Bragg frequency, and the homodyne Lorentzian of double width which is again centered at the origin. This description ignores electronic and other noise, which is assumed to be uncorrelated to the signal and which can, therefore, be simply subtracted after recording and averaging the spectra. The homodyne term is known to be seriously affected by multiple scattering but also by convection or shear [6, 52]. In particular for larger scattering volumes as used here, both result in the loss of coherence of the scattered light, which renders the homodyne term ill-defined. From Eqn. (4), however, we note that the desired information about the diffusive particle motion is also fully contained in each of the super-heterodyne Lorentzians which are symmetric about the origin. In Fig. 2. and also the following, we therefore display the measured data only for positive frequencies centered about the positive Bragg shift frequency.

Eqn. (4) neglects any contribution from non-diffusive particle motions. For a further treatment of this term for flowing systems, see e.g. [42]. Active suspensions are addressed in [41]. Eqn. (4) further ignores any contribution not stemming from single scattering events to the spectra. There are, however, several noise sources including detector shot noise (resulting in a frequency independent background), delta peak type electronic noise and a systematic contribution from the superposition of reference beam light with light scattered off the illuminating beam by parasitic reflection at the cell or IMB surface (resulting in a delta peak at exactly the Bragg frequency).

## Results

Before evaluation, we first mask the data at $\omega_B$ and subtract the frequency independent noise background. We show a background corrected spectrum in Fig. 2a. Still, at 3.2 kHz a $\delta$-type electronic signal of unknown source is observed. Also this is masked. We perform a least square fit of the middle term of Eqn. (4) to the corrected data (confidence level 0.95). Note that here, the measurement duration was 5 min, which was sufficient to obtain a signal to noise ratio of about 10. The fitted expression (red solid line) excellently describes the experimental data. It returns the half width at half maximum, for which we expect HWHM = $2\pi\, q^2\, D_{eff}$ from Eqn. (4). In Fig. 2a, we have HWHM = (75.87±1.3) Hz, where the uncertainty is given as standard error from a fit at confidence level 0.95. This leads to $D_{eff} = D_0 = k_B T / 6\pi\eta a_h$ = (4.06 ± 0.08) $10^{-12}$ m$^2$s$^{-1}$, where the uncertainty now also includes the uncertainties of the temperature reading ($T$ = (298.3 ± 0.8) K (recorded as average of ambient temperature in SH-DLS lab over the day of measurement) and of the viscosity [51]. Fig. 2b shows the q-dependence of the HWHM. The least square quadratic fit (red solid line) demonstrates that this indeed applies. The value for the average effective diffusion coefficient from this fit is $D_{eff}$ = (4.23 ± 0.05) $10^{-12}$ m$^2$s$^{-1}$. Note the absence of any systematic deviations from the expected functionality demonstrating that the suspension indeed is non-interacting.

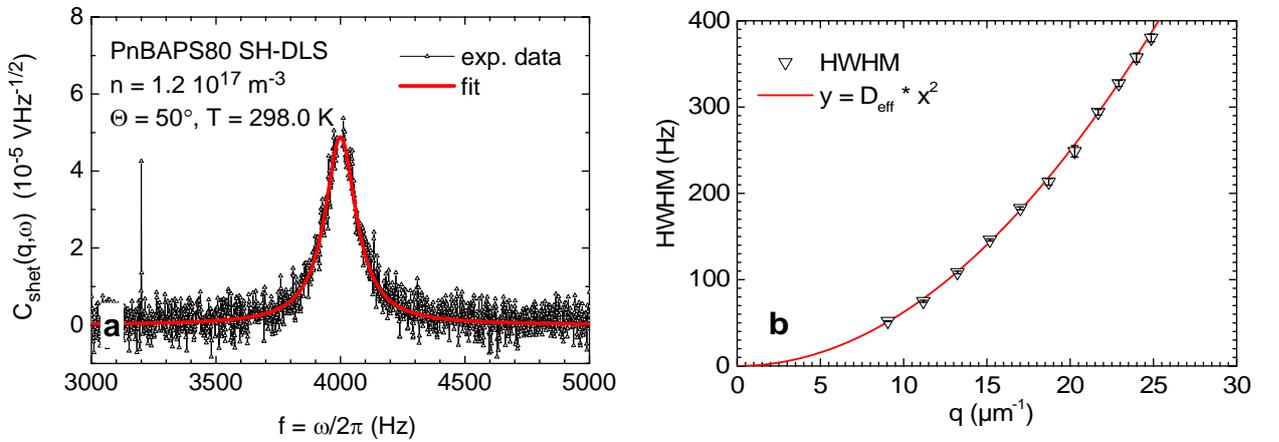

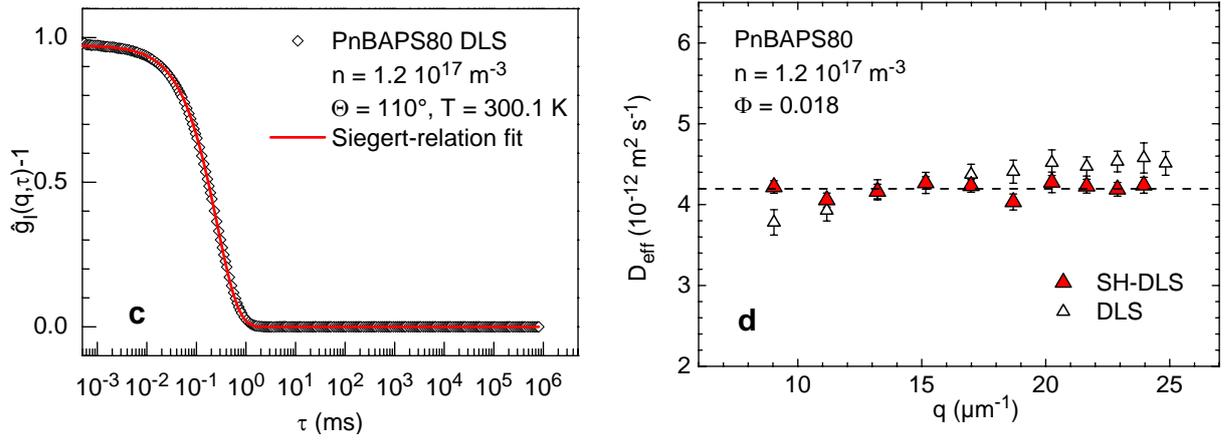

**Fig. 2** Results from SH-DLS and comparison to DLS reference measurements. **a)** Background noise corrected power spectrum of PnBAPS80 at n = 0.12 µm$^{-3}$ taken at $\Theta$ = 50° (q = 11.17 µm$^{-1}$) with fitted Lorentzian (red solid curve). The excellent fit yields $D_{eff}$ = (4.06 ± 0.08) 10$^{-12}$ m$^2$s$^{-1}$. **b)** The half width at half maximum HWHM measured over an angular range between 40° and 140° shows the expected quadratic dependence on $q$. The fit returns $D_{eff}$ = (4.23 ± 0.05) 10$^{-12}$ m$^2$s$^{-1}$. **c)** Reference measurement using homodyne DLS at $q$ = 21.7 µm$^{-1}$. Exploiting the Siegert relation, data are fitted as $\hat{g}_I(q,\tau)$ -1 = $\hat{g}_E(q,\tau)^2$ = $\hat{g}_E(q,0)$ exp(-2$D_{eff}$ $q^2|\tau|$) which returns an intercept of $\hat{g}_E(q,0)$ = 0.972 and a decay constant corresponding to $D_{eff}$ = (4.46 ± 0.07) 10$^{-12}$ m$^2$s$^{-1}$ assuming (298.15±0.3) K. **d)** Wave vector dependent effective diffusion coefficients from SH-DLS (filled symbols) and homodyne DLS (open symbols). The dashed line denotes the average value of the SH-DLS data: <$D_{eff}$> = (4.19 ± 0.06) 10$^{-12}$ m$^2$s$^{-1}$. As expected, this value coincides within statistical uncertainty with the value obtained from the fit in Fig. 2b. The DLS data, however, show a systematic variation attributed to a drift of ambient temperature in the DLS lab.

For comparison we show in Fig. 2c a representative autocorrelation function as obtained in a measurement using homodyne DLS at an angle of $\Theta$ = 110° ($q$ = 21.6 µm$^{-1}$). For a total measurement duration of 120 min the signal to noise ratio of better than 10$^3$ was obtained between 2 and 2 × 10$^4$ ms. For comparison to the SH-DLS data, we first fitted a single exponential to obtain and an intercept of $\hat{g}_E(q,0)$ = 0.972 and a decay constant corresponding to an effective diffusion coefficient of $D_{eff}$ = (4.46 ± 0.07) µm$^2$ s$^{-1}$. We further performed a 2$^{nd}$ order cumulant fit to the same data (not shown), yielding $D_{eff}$ = (4.39 ± 0.04) µm$^2$s$^{-1}$ and PI$_{DLS}$ = 0.035±0.009, which is to be compared to the PI from SAXS of PI$_{SAXS}$ = 0.08. This large discrepancy should be resolvable using non-linear cumulant analysis [38]. In all cases, the quoted uncertainties derive from the standard error for fits performed at a confidence level 0.95 and for an assumed temperature of $T$ = (298.15±0.3) K.

In Fig. 2d, we compare the results of both experiments in terms of $D_{eff}$. The SH-DLS data scatter about a constant average value $<D_{eff}> = (4.19 \pm 0.06)\ 10^{-12}\ m^2s^{-1}$. By comparison, the DLS data show a small, apparently systematic increase at smaller $q$-values. At larger $q$-values, the variation is within statistical accuracy, but the values are above those from SH-DLS. This behavior closely follows the recorded 5.5°C increase of ambient temperature in the DLS lab over the first hours of the measurements performed starting from small $q$ (from some 22°C in the morning to 27.5°C in the afternoon), while the SH-DLS lab stayed at $T = (298.3 \pm 0.8)$ K during the measurements. Using the $D_{eff}(T)$ obtained in DLS at the recorded temperatures, $T$, to calculate the effective diffusion coefficient for $T$ = 298.3K results in values, $D_{eff}(298.3K)$ which coincide with the SH-DLS data within statistical uncertainty (including standard error, uncertainty in temperature reading and resulting uncertainty in viscosity [51], not shown). We therefore conclude, that both experiments are able to measure *average* diffusion coefficients with good accuracy and acceptable precision. For non-interacting suspensions they deliver quantitatively coinciding values independent of scattering angle and we finally obtain an hydrodynamic diameter of $2a_h = (99.6 \pm 2.0)$ nm for PnBAPS80 as averaged over the SH-DLS data and the temperature corrected DLS data. An evaluation and comparison of size dispersities from both experiments has to be postponed to later stages of the project.

## Discussion

We have constructed a prototype version of a SH-DLS instrument capable of performing routine diffusion experiments over a large angular range. The instrument covers the full range of accessible wave vectors known from DLS and in principle already now can be employed for measuring $q$-dependent collective diffusion. Remaining limitations in angular range result from the specific mounting of the sending and detection optics can be overcome by implementing the mounting scheme demonstrated in [17]. The instrument then will also cover the small $q$-range to measure self diffusion from incoherently scattered light. Both the angular resolution and the reproducibility of $q$-adjustment can be improved further by implementing a stepper-motor drive. Variations of detected time averaged intensity could be further reduced replacing standard sample vials by custom made optical cuvettes. Temperature readings can be further improved by directly measuring the IMB fluid temperature.

Already now, however, we could demonstrate the excellent performance of our instrument. The statistical uncertainty (standard error at confidence level 0.95) of an individual measurement is on the order of 1.5%. The performance is therefore fully comparable to that of homodyne time domain DLS. Note, however, that both approaches fall way short in comparison to other, albeit much more time consuming optical approaches, e.g optical tracking. There, for instance, Garbow et al. could resolve all six species of a mixture with sizes ranging between 300 and 450 nm to obtain the average diameters with a combined statistical and systematic relative uncertainty of 0.7% in a single tracking experiment of 21h duration [53]. One further observes that both DLS and SH-DLS radii are consistently larger than the radius derived from static scattering. This, however, is a well-known effect that has been discussed extensively in the literature [8, 54, 55].

As compared to DLS, SH-DLS affords some additional instrumentation. It requires a set of acousto-optical modulators to realize the frequency difference between $I_{ref}$ and $I_{ill}$ and a double-arm goniometer. It is, however, much easier to align: feeding a laser through the detector-sided fibre, the obtained observation beam simply has to be made co-linear with the reference beam by coupling it into the reference beam sending optics fibre coupling, and its lateral extension has to be adjusted to encompass a major central portion of the illumination beam traversing the sample cell by adjusting the distance between the $q$-selecting vertical slit aperture and the detection side fibre coupler grin lens.

The present demonstration is certainly preliminary in the sense that several standard evaluation procedures known from conventional DLS have not yet been implemented and the full special abilities of SH-DLS have not yet been exploited. Performance on ergodic colloidal fluids is expected to be similar to DLS. Samples under flow and active matter, however, can only be studied by SH-DLS, since only there the intermediate scattering function containing the information on both velocity and diffusion is directly accessible without using the Siegert relation [33]. This could e.g. be used to obtain quantitative access to the coupling between structure and diffusivity in shear flow [56].

Even more interesting are measurements in turbid samples. Here, standard DLS is at loss even for weak multiple scattering contributions. In the limit of very strong multiple scattering Diffusive Wave Scattering can be applied, but for all intermediate cases only cross correlation schemes give access to the desired singly scattered light. Here we expect the present goniometer-based SH-DLS to show similar performance as the small angle instrument, where diffusion and directed motion have been studied with a precision of 2% at transmissions as low as 40% [27], while the limiting transmission for signal detection was 20%. Cross correlation experiments clearly performed better in the latter respect, since they detect singly scattered light only and can omit post recording discrimination. In fact, depending on the employed scheme, the detection threshold there is on the order of 1-5% transmission [15, 16, 17, 18, 25]. More important seems to be the precision with which the average relaxation time can be determined. We here obtained $<D_{eff}>$ = (4.19 ± 0.06) $10^{-12}$ $m^2s^{-1}$ from the average of 10 measurements at different angles, each of about five min duration. This is on the order of the precision and duration for the single angle DLS reference measurement, and it can surely be improved further. The initial step will be the inclusion of size dispersity analysis. Here, several approaches are conceivably. First, one may try a Fourier-back transform of the isolated SH contribution of the spectra followed by time domain employing known algorithms. Second, since $\hat{g}_E(q,\tau)$ is the Laplace transform of the normalized distribution of decay rates, one may attempt a deconvolution directly in frequency space. Once that is achieved, one can test the precision obtainable in turbid samples against the current benchmarks set by cross correlation techniques. Moreover then, also the ability to measure q-dependent statics and dynamics in ordered samples can be tested.

Finally, in small angle SH-DLS, we also exploited the excellent ensemble average stemming from the large observation volume to study diffusion in non-ergodic (polycrystalline) materials. This should also be possible with the present instrument, however, now also covering the interesting $q$-region around the structure factor maxima. Less clear and remaining to be tested is the resolution in the case of multiple relaxation times. Preliminary experiments on bimodal mixtures of non-interacting spheres indicate that relaxation times differing by about one order of magnitude can be discriminated. However, the performance in systems showing stretched or compressed exponential, or even more complex relaxation behaviour remains to be explored. The ability to correctly analyse such situations is crucial for the use of SH-DLS in probing dynamics in strongly interacting systems and matching the performance of cross correlation or tracer experiments. Also here, it may turn out very advantageous, that SH-DLS works in the frequency domain, where the different signal contributions are neatly separable. In principle, after correction for multiple scattering, subtraction of background noise, isolation of the SH-signal and shifting in frequency space, the Fourier back transform of the SH part of the spectrum yields the time domain single scattering dynamic structure factor without the use of Siegert's relation. This then can be subjected to the existing evaluation procedures for correlation functions. At present, however, the reliable numerical implementation still presents a formidable challenge.

Concluding, we have taken first steps towards extended diffusion measurements in frequency space reviving early approaches in a goniometer based integral super-heterodyning version. We have demonstrated the excellent performance of the new instrument with simple diffusion measurements and compared to results from conventional homodyne DLS instrumentation. We have discussed remaining weaknesses and open questions as well as outlined the potential scope of our approach. Much work remains to be done. However, we anticipate that after solving the remaining challenges, SH-DLS may become a viable and versatile alternative to access turbid and non-ergodic systems.

## Conflict of Interest

The authors declare that they have no conflict of interest.

[16]